\documentclass[twocolumn, graphics, floatfix, a4paper, aps, prl, longbibliography, superscriptaddress]{revtex4-2} %longbibliography
\usepackage{graphicx}
\usepackage{bm}
\usepackage[usenames,dvipsnames]{xcolor}
\usepackage[colorlinks=true,urlcolor=blue,citecolor=blue,linkcolor=blue]{hyperref}
\usepackage{amssymb,amsmath}
\usepackage[normalem]{ulem}
\usepackage{booktabs}
\usepackage{longtable}
\usepackage{braket}
\usepackage[T1]{fontenc}
\usepackage[utf8]{inputenc}
\usepackage[english]{babel}
\usepackage{xspace}
\usepackage{epstopdf}
\usepackage[final]{microtype}
\usepackage{enumitem}
\usepackage{subfigure}
\usepackage{amsmath}

\newcommand{\qql}{\textquotedblleft}
\newcommand{\qqr}{\textquotedblright\xspace}

\newcommand{\meanv}[1]{\langle #1 \rangle}

\graphicspath{{./Figures/}}
\begin{document}
\title{Suppression of non-equilibrium quasiparticle transport in flat band superconductors}

\author{Ville A. J. Pyykk\"onen}
\affiliation{Department of Applied Physics, Aalto University School of Science,
FI-00076 Aalto, Finland}

\author{Sebastiano Peotta} 
\affiliation{Department of Applied Physics, Aalto University School of Science, FI-00076 Aalto, Finland}

\author{P\"aivi T\"orm\"a}
\email{paivi.torma@aalto.fi}
\affiliation{Department of Applied Physics, Aalto University School of Science, FI-00076 Aalto, Finland}

\begin{abstract}
We study non-equilibrium transport through a superconducting flat-band lattice in a two-terminal setup with the Schwinger-Keldysh method. We find that quasiparticle transport is suppressed and coherent pair transport dominates. For superconducting leads, the AC supercurrent overcomes the DC current which relies on multiple Andreev reflections.  With normal-normal and normal-superconducting leads, the Andreev reflection and normal currents vanish. Flat band superconductivity is thus promising not only for high critical temperatures but also for suppressing unwanted quasiparticle processes. 

%We study non-equilibrium transport through a piece of flat-band lattice system acting as the scatterer in a two-terminal setup. We find that the quasiparticle transport is quenched through a flat band partially filled with attractively interacting particles, with only a small or vanishing contribution depending on the model. In practice, normal and Andreev reflection currents in normal-normal and normal-superconducting junctions are small even at finite interactions. Also, the multiple Andreev reflections are not present in a superconducting contact. The main transport channel is the pair transport: supercurrent in a superconducting setup is important for attractively interacting particles. We mainly consider the transport of coherent pairs, that is, the Cooper pairs due to the limitations of the mean-field approach. Nevertheless, incoherent pair transport is also a possibility. We give some general estimates on the importance of these 'preformed pairs' as a transport channel. The quenching of the dissipative quasiparticle processes presents a potential solution to the quasiparticle poisoning problem present in e.g. superconducting qubits. Also, lack of dc current out-of-equilibrium means that the system acts as a pure dc to ac converter. We briefly discuss the practical implications of the found phenomena.
\end{abstract}

\maketitle

The destructive interference of waves scattered from a periodic potential can lead to disorder-free localization, which manifests as the vanishing of the energy width of a band in the band structure~\cite{leykam2018}. Currently, an important goal is to engineer materials in which these so-called flat bands occur, as this generally allows to enter a strongly correlated regime with emerging exotic phases. Well known instances of this general picture are the fractional quantum Hall effect and Chern insulators~\cite{parameswaran2013,bergholtz2013}, and more recently twisted bilayer graphene and similar moir\'e materials~\cite{cao2018,macdonald2019,andrei2020,balents2020,kennes2021,andrei2021}, which are built by stacking and twisting atomic layers with various compositions.

The discovery of superconductivity at the \qql magic\qqr angle in twisted bilayer graphene (TBG) has amplified the interest in the problem of superconductivity in the flat band limit and its competition with other phases, such as correlated insulators~\cite{cao2018,balents2020,torma2022}. This is a challenging problem due to strong correlations, nevertheless it has received a lot of attention~\cite{peotta2015,julku2016,liang2017,torma2018,xie2020-2,hu2019,julku2020,iskin2019,iskin2021,peri2021,herzog-arbeitman2022-1,pyykkonen2021,huhtinen2022,chan2022,chan2022-1,kitamura2022,hofmann2020,wang2020-1,orso2022} and even exact results can be derived for the ground state and excitations under some conditions~\cite{tovmasyan2016,herzog-arbeitman2022}, moreover, it was shown that superconductivity in flat bands originates from quantum geometry and topology~\cite{peotta2015,liang2017,xie2020-2,huhtinen2022}. Flat band superconductivity is particularly promising as a route for higher temperature superconductivity, as the critical temperature is linearly proportional to the interaction energy~\cite{heikkila2011,kopnin2011,khodel1994}, while it is exponentially suppressed for weak-coupling in the case of a dispersive band~\cite{parravicini2013}. 

%PT: I made an alternative to this paragraph
%An open interesting question is what are the transport properties of the superconducting state in the flat band limit: they are expected to be rather unusual due to the coexistence of localized (or almost localized) fermionic quasiparticles and mobile bosonic Cooper pairs~\cite{tovmasyan2018}. For instance, for the sawtooth lattice with a flat band, the two-body problem in the flat band limit \cite{torma2018} gives an effective pair mass at least an order of magnitude smaller than the quasiparticle mass. Fermionic quasiparticles are responsible for dissipative transport in superconductors and superconducting weak-links~\cite{likharev2022}, but it is unclear how this can be the case in a flat band due to the quenching of the kinetic energy. 

A major open question is the transport properties of a superconducting state in the flat band limit. Transport in superconductors and superconductive weak links typically includes non-dissipative AC and DC supercurrents, as well as dissipative transport involving fermionic quasiparticles~\cite{likharev2022}. It has been theoretically shown that equilibrium DC supercurrents are possible in flat bands~\cite{peotta2015,torma2022,pyykkonen2021}, but otherwise little is known about transport. In certain highly symmetric flat-band systems single particles remain localized due to local conserved quantities~\cite{tovmasyan2018}, and quasiparticle excitations have a flat dispersion~\cite{herzog-arbeitman2022}, while pairs can be mobile. These equilibrium results on infinite bulk systems hint that flat band transport could show unique features also in out-of-equilibrium situations, i.e.~under voltage or current bias, and in the presence of interfaces. In this work, we focus on \textit{out of equilibrium}  transport in a lattice model with a flat band in which superconductivity arises due to a local attractive interaction. We find that the nondissipative supercurrent, the current carried coherently by highly mobile Cooper pairs, dominates over the dissipative current involving quasiparticles. The absence of quasiparticle transport and dissipation suggests flat band superconductors as remarkably promising building blocks for quantum devices. 

\begin{figure}
    \centering
    \includegraphics[width = \columnwidth]{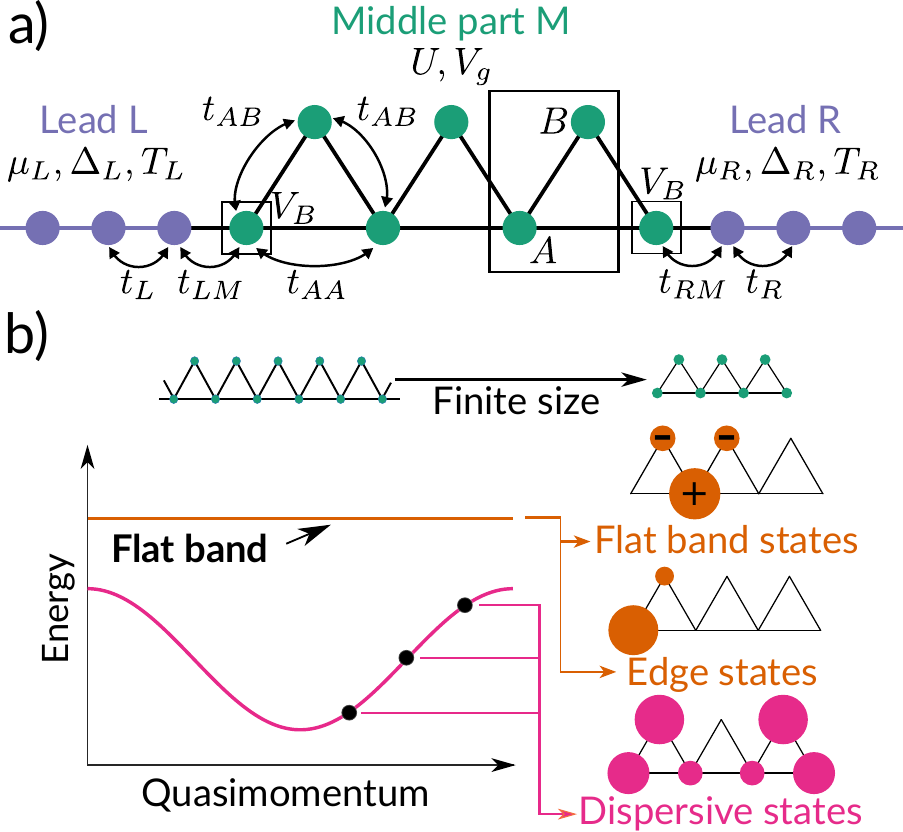}
    \caption{The model used in the calculations. a) The two-terminal setup with a piece of sawtooth lattice (middle part M) connected to two leads modeled by semi-infinite chains, labeled left (L) and right (R). The figure also highlights the sawtooth lattice unit cell which has two sites, labeled A and B, and the edge sites. The graph corresponds to the tight-binding model hopping amplitudes. Here $U<0,  V_g,V_B$ correspond to interaction strength, gate potential and boundary potential in the middle part, and $\mu_{L,R}$, $\Delta_{L/R}$, $T_{L/R}$ are the chemical potential, superconducting order parameter, and temperature of the left and right leads, respectively. (b) Correspondence between the sawtooth lattice and the truncated piece. The flat band states of the infinite system are also present in the finite size system since the destructive interference remains. At the edges, the flat band states are changed to edge states, which we tune to degeneracy with the flat band states by a boundary potential. The dispersive band corresponds to the dispersive states.}
    \label{fig:figure1}
\end{figure}

We address the non-equilibrium transport properties by using the two terminal setup depicted in Fig.~\ref{fig:figure1}~a), where also the notation and the model flat band system, the sawtooth lattice, are presented. In the setup, a middle structure (M) is connected via leads to two reservoirs, left (L) and right (R), respectively. The leads can be either normally conducting (N) or superconducting (S) enabling three different lead configurations: normal-normal (NN), normal-superconducting (NS) and superconducting-superconducting (SS).

The setup is modeled using the following tight-binding Hamiltonian with an Hubbard interaction term
\begin{equation}
    \hat{H} = \sum_{\alpha i\beta j,\sigma}T_{\alpha i,\beta j}\hat{c}_{\alpha i \sigma}^\dagger\hat{c}_{\beta j \sigma} + \sum_{\alpha i} U_{\alpha i} \hat{c}_{\alpha i \uparrow}^\dagger \hat{c}_{\alpha i \downarrow}^\dagger \hat{c}_{\alpha i\downarrow}\hat{c}_{\alpha i\uparrow}~, 
\end{equation}
where $T_{\alpha i, \beta j}$ is the single particle Hamiltonian with $\alpha,\beta \in \{L,R,M\}$ labeling different parts of the system, $i,j$ are site indices and $\sigma = \{\uparrow,\downarrow\}$ is the spin index, and $U_{\alpha i} \leq 0$ is the attractive interaction strength.  The graph in Fig.~\ref{fig:figure1}~a) shows the single-particle tight-binding parameters of the model. Specifically, the single-particle Hamiltonian can be divided as 
$\hat{H}_0 = \hat{H}_{L}+\hat{H}_R+\hat{H}_M +\hat{H}_{\mathrm {contact}}~,$
where $\hat{H}_{L},\hat{H}_R$ and $\hat{H}_M$ are the Hamiltonians of the left lead  (L), right lead (R), and the middle part (M) in between the leads, respectively, and $\hat{H}_{\mathrm{contact}}$ connects them together.
The tight-binding matrices corresponding to the lead Hamiltonians $\hat{H}_{L}$ and $\hat{H}_R$ are given by
$ T_{L/Ri,L/Rj} = -\delta_{ij}\mu_{L/R} +  \delta_{j,i\pm 1} t_{L/R}~,$
where $\mu_{L/R}$ are the chemical potentials of the respective leads
and $t_{L/R}$ are their hopping amplitudes.
The tight-binding matrix related to the middle part Hamiltonian $\hat{H}_{M}$ is given by $T_{Mi,Mj} = \delta_{ij}(-V_g+V_B \delta_{j,\mathrm{edge}}) + t_{M,ij} ~$
where $V_g$ is the gate potential used to control the filling of the middle part states, $t_{M,ij}$ is the model-specific hopping matrix given by the graph in Fig.\ref{fig:figure1}~a), and $V_B$ is a boundary potential at the edge sites used to control the edge state energies \cite{pyykkonen2021}.
The contact Hamiltonian $\hat{H}_{\mathrm{contact}}$ corresponds to $T_{Mi,L/R j} = T_{L/R j,Mi} = t_{L/R,M}\delta_{i,{\rm edge}} \delta_{j,0}~,$ $t_{L/R,M}$ being the respective hopping amplitudes. 

As a specific example of a system with a flat band, we look at the sawtooth ladder shown in Fig.~\ref{fig:figure1}~a). When the hopping amplitudes satisfy the condition  $t_{AB} = \sqrt{2}t_{AA}$, the upper band becomes flat, as shown in Fig.~\ref{fig:figure1}~b), since it is composed of localized V-shaped states. We look at a finite segment of the sawtooth ladder with $N$ unit cells and an additional A site. An earlier study \cite{pyykkonen2021} has shown that flat band equilibrium transport, i.e.~the DC Josephson effect, is possible through a finite segment of the sawtooth ladder while no current is seen in the flat band in absence of interactions. As indicated in Fig.~\ref{fig:figure1}~b), the system has $N-1$ flat band states, $N$ dispersive band states and two edge states exponentially localized to the edges, which result from the edge flat band states due to the absence of one of the B sites. The edge states can be made degenerate with the flat band states by setting the boundary potential at the edge sites to $V_B = -t_{AA}$. This has also the effect of perfectly localizing the edge states to the edge A and B sites \cite{pyykkonen2021}.

We enter the out of equilibrium regime by applying a chemical potential bias $V = \mu_L-\mu_R$. The current is computed with the non-equilibrium Green's functions (NEGF) \cite{haug2008, stefanucci2013} method, also known as Schwinger-Keldysh \cite{schwinger1961,keldysh1965} or Kadanoff-Baym \cite{kadanoff1962} method.  We evaluate the current at the left lead $I_L = 4\mathrm{Im}(t_{ML} \meanv{\hat{c}_{M,\mathrm{left edge}\uparrow}^\dagger\hat{c}_{L,0\uparrow}}),$ which takes into account also the down-spin current by an additional factor of two. The details of the method are given in the Supplementary material \cite{[{See Supplemental Material at }][{ for more details.}]supp}. Similar approaches have been used to study for instance point contacts~\cite{cuevas1996}, quantum dots~\cite{martin-rodero2011}, magic-angle TBG~\cite{alvarado2021}, and many other systems~\cite{jauho1994} at a two-terminal setup. The filling of the middle part states is controlled by the gate potential $V_g$. We limit our attention to the stationary state solutions, where we assume that the initial correlations and the transient effects have vanished. This results in a time-independent solution within the NN and NS junctions and in time-periodic behavior with the SS junctions.

We treat the Hubbard interaction with a self-consistent mean-field approximation: we take both the superconducting order parameter and the Hartree potential into account. The mean-field approximation has been shown to be an accurate description of flat band superconductivity at equilibrium in a number of works \cite{julku2016,tovmasyan2016,liang2017,chan2022,
chan2022-1}.
The mean-field Hamiltonian is compactly written using the Nambu spinors $\hat{d}_{\alpha i} = ( \hat{c}_{\alpha i \uparrow}, \hat{c}_{\alpha i \downarrow}^\dagger)^T$ as
\begin{equation}
\begin{split}
    &\hat{H}_{MF}(t)  = 
    \sum_{\alpha i, \beta j}\\ 
    &\hat{d}_{\alpha i}^\dagger
    \begin{pmatrix}
        T_{\alpha i,\beta j} + V_{H,\alpha i}(t)\delta_{\alpha i,\beta j} & \Delta_{\alpha i}(t)\delta_{\alpha i,\beta j}\\
        \Delta_{\alpha i}(t)^*\delta_{\alpha i,\beta j} &
        -T_{\alpha i,\beta j}^* - V_{H,\alpha i}(t)\delta_{\alpha i,\beta j}
    \end{pmatrix}
    \hat{d}_{\beta j}~,
    \end{split}
\end{equation}
where $\Delta_{\alpha i}(t)$ and $V_{H,\alpha i}(t)$ are the superconducting order parameter and the Hartree potential, respectively, which are determined self-consistently for the middle part utilizing the equations $\Delta_{\alpha i } = U_{\alpha i} \meanv{\hat{c}_{\alpha i \downarrow} \hat{c}_{\alpha i \uparrow}}$ and $V_{H,\alpha i} = U_{\alpha i} \meanv{\hat{c}_{\alpha i \uparrow}^\dagger \hat{c}_{\alpha i \uparrow}}~.$ The leads are considered with a constant, uniform order parameter and their respective Hartree potentials are absorbed into their chemical potentials $\mu_{L/R}$. As the notation reminds, the time-independent Hubbard interaction may result in a time-dependent mean-field theory at non-equilibrium conditions, which is the case now for the SS junctions even in the stationary state.
In the time-periodic situation, we include the harmonics coefficients until they are within the self-consistent accuracy. In addition, we make frequency cutoffs to make the calculation feasible. The self-consistency is determined using the relative maximum error metric and the accuracy we demand is $10^{-5}.$ The flat band filling is sensitive to minute changes in the Hartree potential introducing difficulty in convergence. However, with suitable methods (see Supplementary \cite{[{See Supplemental Material at }][{ for more details.}]supp}) we obtain convergence.

The chemical potential bias $V$ between the leads induces a particle current through the middle part. The current-carrying processes can be classified into two categories within the limits of the mean-field theory: pure coherent pair transport, and processes involving quasiparticle transport. 
Incoherent pair transport and processes involving more complicated $n$-body states are not included in our approach. 
With one or both of the leads being normal, only the quasiparticle-related processes are possible but with the superconducting leads also the coherent pair transport may contribute.

Quasiparticle transport can occur through a channel by a combination of direct transmissions, branch-crossing transmissions, reflections, and Andreev reflections (AR) \cite{blonder1982, datta1996}. The case of normal-normal (NN) reservoir configuration is the simplest, since only direct transmission and reflection are allowed. Instead, in the case of the NS and SS junctions, since there are no quasiparticle states available within the superconducting gap, an AR may occur~\cite{andreev1964} where a particle is reflected as a hole of the opposite spin. In the process, a Cooper pair is transmitted to the superconducting reservoir (or removed in the opposite process).  In the case of an SS junction, quasiparticle transport for a bias smaller than the superconducting gaps is enabled by the multiple AR (MAR), where AR occurs multiple times between the leads until the quasiparticle escapes \cite{klapwijk1982,blonder1982}. Also, the branch-crossing transmission from a quasiparticle into its time-reversed counterpart on the other side may occur at a bias larger than the superconducting gap, however, these are not important in our work.

Coherent Cooper pair transport, that is the Josephson effect, occurs between superconducting reservoirs with relative superconducting order parameter phase difference \cite{josephson1962,likharev2022}. 
%More generally, Cooper pair current is related to a gradient of the order parameter phase. 
Importantly, the order parameter has the time-dependence $\Delta = |\Delta|\exp(i(\phi_0+2E_{F}t))$,  where $E_F$ is the Fermi energy \cite{datta1996}. Therefore, with a constant bias $V$, the superconducting phase difference $\phi$ evolves over time $\phi(t) =\phi_0 + 2Vt$ , leading to an alternating current, i.e.~the AC Josephson effect. In other words, two superconductors with a relative bias is an inherently time-dependent system having no time-independent steady-state solution.

The AC Josephson effect is a coherent and non-dissipative phenomenon based on Cooper pairs only. The AR and MAR are coherent processes that involve both Cooper pairs and quasiparticles, and due to the latter, are dissipative. For the sake of brevity, in the following we refer the AC Josehpson effect as coherent pair transport/process, and AR/MAR as quasiparticle transport/process.

\begin{figure}
    \centering
    \includegraphics[width=1.00\columnwidth]{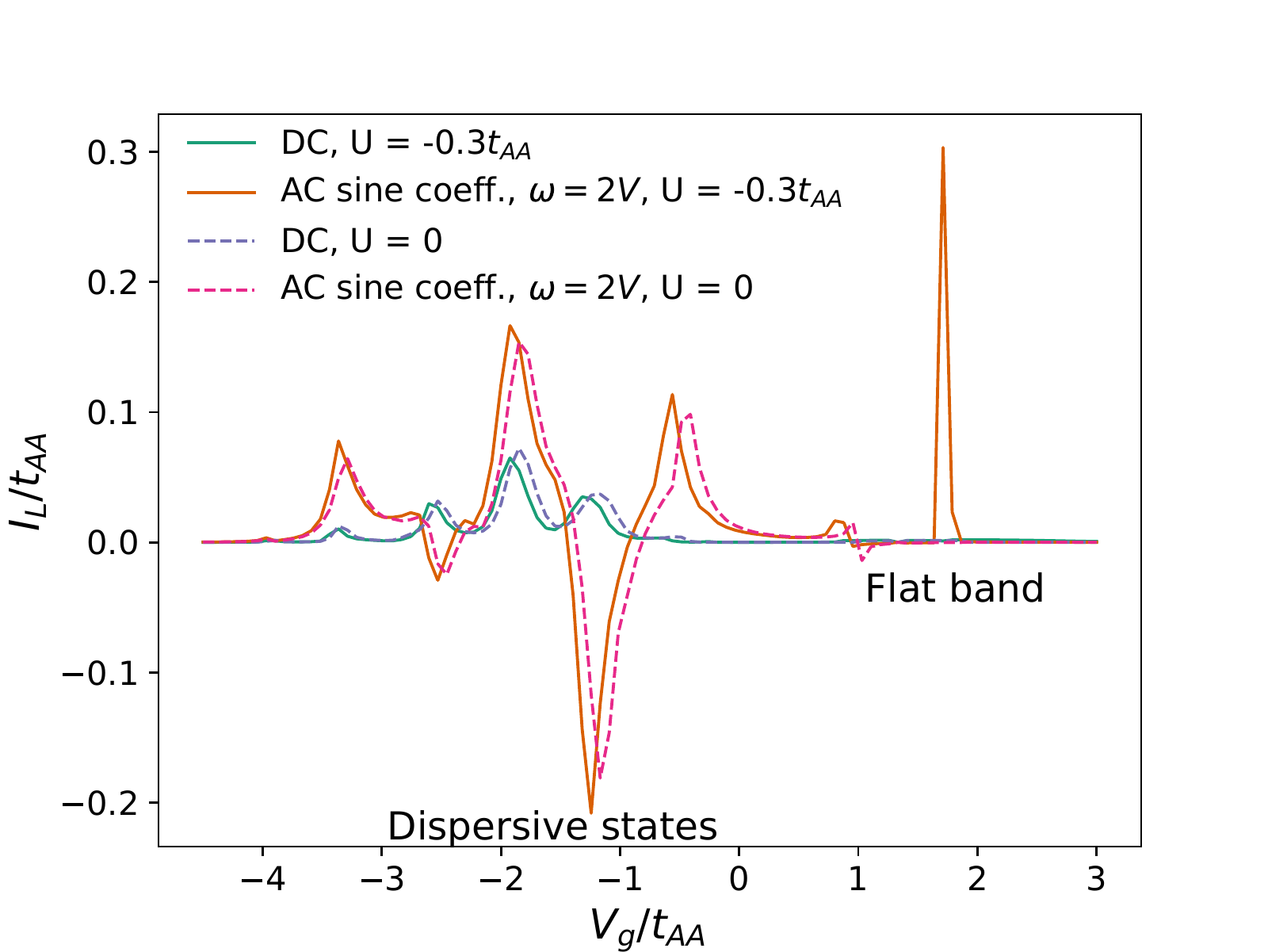}
    \caption{Left lead current $I_L$ through the sawtooth lattice with three unit cells and an additional site between two superconducting leads at a varying gate potential $V_g$ with a constant bias $V$. The parameters are $U = -0.3 t_{AA}$, $V = 0.5 t_{AA}$, $t_{LS} = t_{RS} = 5.3 t_{AA}$, $\Delta_{L/R} = -t_{AA}$, $T_{L/R}= 0$, and the leads are in the wide-band limit with $t_L = 30 t_{AA}.$ The DC current is dominated by the quasiparticle MAR processes, which are finite around the dispersive states located at gate potential $-4 < V_g/t_{AA} < 0$, where there are current peaks. The quasiparticle current around the flat band, located around $V_g/t_{AA}= 2$, vanishes. In strong contrast, the AC Josephson current, that is, the AC sine component of the current at the Josephson frequency $2V$ has in addition to the peaks at the dispersive states a prominent peak corresponding to the flat band states.}
    \label{fig:figure2}
\end{figure}

Firstly, we look at transport through an SS junction since it provides the clearest connection to the known equilibrium transport features with a similar setup, presented in Ref. \cite{pyykkonen2021}. It also allows a direct comparison between the pair and the quasiparticle contributions, namely the AC Josephson effect and the MAR. As mentioned above, the stationary solution of an SS junction at a time-constant bias $V$ is time-periodic with the period of $\tau = \pi/V$ (in units where $\hbar  = e = 1$). Fig.~\ref{fig:figure2} presents the DC component and the first harmonic AC sine component of the current through the sawtooth lattice at a constant bias $V$ and varying gate potential $V_g$, which controls the filling. The two flat band states and the two edge states lie at the gate potential $V_g = -2 t_{AA}$ and the three dispersive band states are between $V_g = 4t_{AA}$ and $V_g = 0.$ The states corresponding to the dispersive band exhibit a finite AC current, where the amplitude variation shows Fano-resonance type behavior. The DC component, which corresponds to quasiparticle MAR processes, exhibits current peaks related to the dispersive states. There are more peaks than the corresponding three dispersive states: MAR depends on the particular path in energy a quasiparticle passes \cite{martin-rodero2011}, resulting in sensitivity to the gate potential $V_g$ and many local maxima. In general, the dispersive band acts as a point-contact channel in an expected manner, and having interactions in the middle part ($U\neq 0$) has no qualitative effect. The flat band states, in contrast, have no current in the zero interaction case, but the AC sine component has a large amplitude in presence of interactions (and superconductivity) in the middle part. Most remarkably, in strong contrast to the AC component, the flat band DC current vanishes even at finite interaction. This indicates that the quasiparticle transport is quenched. 

% \begin{figure*}
%     \centering
%     \includegraphics[width=0.6\columnwidth]{Figures/Figure2.eps}
%     \caption{Comparison. Parameters (when applicable) three unit cells and one site, $t_{AB} = \sqrt{2}t_{AA} = -\sqrt{2}$, $t_{LS} = t_{RS} = 5.3t_{AA}$, $V=0.33t_{AA}$ $\Delta_L,\Delta_R = t_{AA}, U = -t_{AA}/2$,$t_{L}=t_{R} = 30 t_{AA}$, $V_{C,L} =V_{C,R} = -t_{AA}, T_L = T_R = 0.$ Flat band states are at gate $V_g = 2$. The current for NN and NS junctions and the DC current at SS junctions are small. The AC Josephson current amplitude is large.}
%     \label{fig:figure2}
% \end{figure*}
% \textbf{Fifth paragraph: suppression of quasiparticle current: NN and NS}
% \begin{itemize}[noitemsep]
%     \item NN and NS results for sawtooth and Creutz ladder 

% \end{itemize}
% \begin{figure}
%      \centering
%      \subfigure{
%          \centering
%          \includegraphics[width=\textwidth]{figure2a}
%          \caption{Biased superconducting junction}
%          \label{fig:ss_ne}
%      \end{subfigure}
%      \hfill
%      \begin{subfigure}[b]{0.8\columnwidth}
%          \centering
%          \includegraphics[width=\textwidth]{figure2b}
%          \caption{Normal-Normal and Normal-Superconducting junctions}
%          \label{fig:nx_ne}
%      \end{subfigure}
%         \caption{Current through sawtooth lattice with 3 unitcells and a site with varying gate at constant bias.}
%         \label{fig:Figure2}
% \end{figure}

\begin{figure}
    \centering
    \includegraphics[width=1.00\columnwidth]{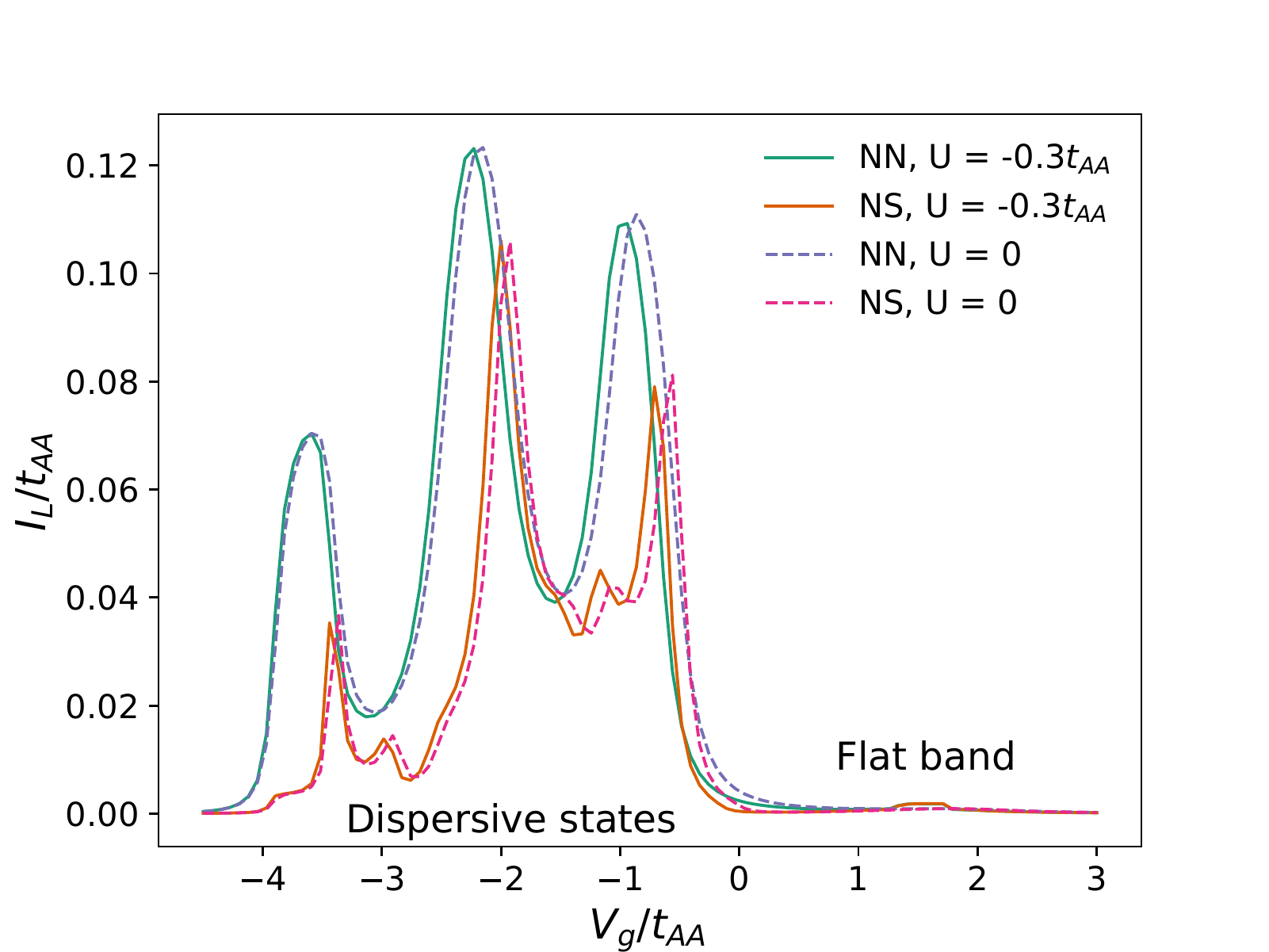}
    \caption{Left lead current $I_L$ through the sawtooth lattice with three unit cells and an additional site in an NN and NS lead configuration at a varying gate potential $V_g$ with a constant bias $V$. The dispersive states correspond to peaks in the current. Flat band states correspond to only very small peaks, in presence of interactions, in both cases. The parameters are otherwise the same as in Fig.~\ref{fig:figure2} but the order parameters are zero for the normal leads. Coherent pair transport does not contribute to the transport: in the NS configuration, the Andreev reflection is the only means for transport since the bias $V$ is smaller than the supercondcuting order parameter amplitude $\Delta_R$, whereas in the NN case the usual transmission and reflection govern the transport. Even though the middle part acquires a finite order parameter due to proximity effect, the superconducting order parameter phase is constant at the flat band energy and thus there is no coherent pair current. The small flat band current contribution is due to the Hartree potential inhomogeneity. }
    \label{fig:figure3}
\end{figure}

We have also considered the NN and NS lead configuration transport through a flat band. The results are shown in Fig.~\ref{fig:figure3}. The dispersive states in the NN case correspond to clearly defined peaks when their filling is varied by the gate potential $V_g$. The Andreev reflection in the NS case varies less smoothly, with sharper peaks around the dispersive region. However, the flat band current is small with both configurations, similarly to the MAR-dominated DC transport in the SS junction. The current in the NN and NS cases is related to quasiparticle current as, in terms of the mean-field theory, coherent pair current would require a superconducting order parameter phase gradient. In the case of the NN junction, the lack of pair transport processes is clear since it turns out that the self-consistent order parameter at the sawtooth lattice vanishes.  In the NS configuration, the phase of the order parameter there is found to be uniform even if the amplitude varies. Thus, there is no coherent pair current. The small flat band current is caused by the Hartree potential inhomogeneity, which affects the destructive interference that causes localization, thereby enabling a quasiparticle current. It is remarkable that this Hartree-potential-induced quasiparticle current remains small, even at non-equilibrium. 

%PT: I made an alternative to this
%Incoherent pair transport is beyond the mean-field picture of superconductivity so it is difficult to address it directly. However, we may try some guesses to assess its contribution to the transport. In case of some flat band models, such as the Creutz ladder, it has been shown that the preformed pairs would be the only possible means for transport above $T_c$ \cite{tovmasyan2018}.
%In the sawtooth lattice, such argument is not valid since the unit cell sites are not symmetric: they may have different densities leading to effective disruption of the flat band condition.
%However, we have found that, at the flat band in the sawtooth ladder, the mean-field quasiparticle mass is over an order of magnitude larger than the bound pair effective mass. Thus, the incoherent pair transport could be significant in principle. 
%In order to capture it theoretically, the theory would need to be expanded to take into account effective non-local interactions e.g. by the second Born approximation including irreducible Feynman diagrams of the second order in the interaction. 

Incoherent pair transport is not included in our mean-field theory approach. However, based on standard knowledge of superconducting junctions, it should be small. In a usual SS junction with a barrier between the superconductors characterized by a tunneling coupling $t$, the coherent pair transport (Josephson effect) is proportional to $t^2$ while incoherent pair-tunneling is a process of the order $t^4$, which for $t<1$ is suppressed \cite{mahan2000}.  In our case, the inverse of the pair mass in the middle flat band region gives effectively the tunneling coupling between the superconducting leads. By the analogy, it is then justified to first consider only the coherent pair transport. However, as flat bands have shown many surprises, it is worthwhile to study in the future whether incoherent pair transport could be relevant, against expectations. Even if it is, the suppression of fermionic quasiparticle transport discovered here will have consequences to transport and device properties.

The system of Fig.~\ref{fig:figure1} can be realized in ultracold gas two-terminal setups~\cite{pyykkonen2021}. Our predictions can be tested there once low enough temperatures are reached to make the middle part superconducting, in addition to the leads \cite{husmann2015,huang2022}. One could also amend other ultracold gas platforms that demonstrated the Josephson effect~\cite{albiez2005,valtolina2015,luick2020}. In TBG devices, Josephson junctions can be defined by gate configurations on a single graphene bilayer \cite{de_vries2021,rodan-legrain2021,diez-merida2021}. For instance SIS, SS’S, and SIS’IS junctions have been realized, where "S" is one type of superconductor and ''S’'' another, and ''I'' denotes an insulating state. Close to our scenario are the two last ones where S would be a superconductor in a part of the band that is maximally dispersive, and S’ an area where the Fermi level is gated to match the flattest part of the band. One could also realize the desired junction by TBG material between superconducting leads.  

In summary, we have shown by non-equilibrium self-consistent mean-field theory that coherent pair transport, i.e.~the AC Josephson effect, largely dominates the flat band transport in a saw-tooth ladder connected to superconducting leads. The transport in the dispersive band shows the usual behaviour: AC Josephson current, as well as DC current realized via MAR. In contrast, quasiparticle transport is prominently absent in the flat band case. With the leads in the NN and NS configurations, there is no current through the flat band, even when a small proximity-induced order parameter exists in the middle part. Again, this means that quasiparticle processes such as Andreev reflection are suppressed in the flat band. 
Equilibrium studies suggested quasiparticle localization for uniform systems of high symmetry~\cite{tovmasyan2018,herzog-arbeitman2022}. It is remarkable that this is the case also in a \textit{non-equilibrium} setting, with interfaces that could lead to complex processes akin to AR and MAR, and in a lattice where Hartree effects combined with reduced symmetry may also induce transport. 

So far, the main motivations of flat band superconductivity studies have been high critical temperatures and strong correlations. Our results highlight that their \textit{unique transport properties} make flat band superconductors promising for quantum devices. In usual superconductors, there is always dissipation associated with AC currents at finite temperature due to quasiparticle current, termed as normal current in two-fluid models~\cite{tinkham2004}. This dissipation is present even at low frequencies and grows as frequency squared, limiting high-frequency operation. An intriguing prospect, which our results promote, is that a flat band superconductor where quasiparticle transport is quenched could enable \textit{ultra-low dissipation (low-power), high-frequency superconducting AC devices}. Flat band superconductors might offer a “cure for quasiparticle poisoning”; quasiparticles, numerous at non-equilibrium even at low temperatures, limit the coherence of quantum bits based on Josephson junctions \cite{aumentado2004,catelani2022} and Majorana nanowires \cite{rainis2012,higginbotham2015}, and lower the sensitivity of kinetic inductance detectors \cite{de_visser2012}. Our results suggest that a flat-band-superconductor part in a device would block the transport of quasiparticles, even at a non-equilibrium situation, while letting supercurrent through. %de_visser2011

\begin{acknowledgments}
P. T. and V. P. acknowledge support by the Academy of Finland under project numbers
307419, 327293, 349313.
V. P. acknowledges financial support by the Jenny and Antti Wihuri Foundation. 
S. P. acknowledges support
from the Academy of Finland under Grants No. 330384
and No. 336369.
\end{acknowledgments}

\end{document}

\title{Suppression of non-equilibrium quasiparticle transport in flat band superconductors -- Supplemental Material}

\author{Ville A. J. Pyykk\"onen}
\affiliation{Department of Applied Physics, Aalto University School of Science,
FI-00076 Aalto, Finland}

\author{Sebastiano Peotta} 
\affiliation{Department of Applied Physics, Aalto University School of Science, FI-00076 Aalto, Finland}

\author{P\"aivi T\"orm\"a}
\email{paivi.torma@aalto.fi}
\affiliation{Department of Applied Physics, Aalto University School of Science, FI-00076 Aalto, Finland}

\maketitle
\tableofcontents

\section{Details of the model}
\subsection{Mean-field Hamiltonian for the two-terminal setup}
We consider the two-terminal device, illustrated in the main text Fig.~1, with two leads connected to a 
lattice structure described by the mean-field Hamiltonian
\begin{equation}
    \hat{H} = \hat{H}_L +\hat{H}_R +\hat{H}_M + \hat{H}_{\mathrm{contact}}~,
\end{equation}
where the terms are Hamiltonians on left lead, right lead, middle structure and the hopping between the leads and the structure, respectively.
They are defined by
\begin{gather}
    \hat{H}_L =\sum_{ij\sigma} (-\mu_L\delta_{ij}+ t_L\delta_{j,i\pm 1})\hat{c}_{L,i\sigma}^\dagger \hat{c}_{L,j\sigma} + \sum_i \left(
    \Delta_L \hat{c}^\dagger_{L,i\uparrow}\hat{c}^\dagger_{L,i\downarrow}
    +h.c. \right)~,\\
    \hat{H}_R = \sum_{ij\sigma} (-\mu_R\delta_{ij}+ t_R\delta_{j,i\pm 1})\hat{c}_{R,i\sigma}^\dagger \hat{c}_{R,j\sigma} + \sum_i \left(
    \Delta_R \hat{c}^\dagger_{R,i\uparrow}\hat{c}^\dagger_{R,i\downarrow}
    +h.c. \right)~,
\end{gather}
where $\mu_{L/R}$ are the chemical potentials, $t_{L/R}$ are the hopping amplitudes and $\Delta_{L/R}$ are the superconducting order parameters of the leads. The Hamiltonian of the middle structure is given by 
\begin{equation}
    \begin{split}
    \hat{H}_M =& \sum_{ij,\sigma} ([-V_g+V_{H,i\sigma}+V_B\delta_{i,\mathrm{edge}}]\delta_{ij}+t_{M,ij})
    \hat{c}^\dagger_{M,i\sigma}\hat{c}_{M,j\sigma} \\
    +& \sum_i \left(
    \Delta_{M,i}\hat{c}^\dagger_{R,i\uparrow}\hat{c}^\dagger_{R,i\downarrow}
    +h.c. \right)~,
    \end{split}
\end{equation}
where $V_g$ is the gate potential, $V_B$ is the boundary potential utilized to tune the edge states, $t_{M,ij}$ is the hopping matrix determining the lattice geometry. Furthermore, $V_H,\Delta_{M,i}$ are self-consistent Hartree potential and superconducting order parameter, respectively, solved conjointly with the equations $V_{H,i\sigma} = U\meanv{\hat{c}^\dagger_{M,i\bar{\sigma}}\hat{c}_{M,i\bar{\sigma}}}$, where $\bar{\sigma}$ denotes the opposite spin to $\sigma$, and $\Delta_{M,i} = U \meanv{\hat{c}_{M,i\downarrow}\hat{c}_{M,i\uparrow}}$. 
Finally, the contact Hamiltonian is
\begin{equation}
    \hat{H}_{\mathrm{contact}} = \sum_{\sigma} \left(
    t_{LS}\hat{c}^\dagger_{L,0\sigma}\hat{c}_{M,c_{L}\sigma}
    +t_{RS}\hat{c}^\dagger_{R,0\sigma}\hat{c}_{M,c_{R}\sigma} +h.c.\right) ,
\end{equation}
where $t_{L/R,S}$ are the hopping amplitudes between the leads and the system and $c_{L/R}$ labels the contact point of the leads on the middle structure.

Despite its appearance, the  mean-field solution is not time-independent. For bulk superconductors, such as the leads, it holds that $\Delta = |\Delta|\exp(i2\mu t)$, where $\mu$ is its chemical potential \cite{datta1996}. It directly follows that at a two-terminal system with two superconducting leads, the difference in the frequencies, $2V\equiv 2\mu_L-2\mu_R$ introduces time-periodicity to the Hamiltonian. To simplify the calculations, we make a canonical transformation into a basis where the chemical potential is zero but the hoppings are time-dependent. In other words, the local energy is measured with respect to the local chemical potential. This is obtained by defining $\ket{\psi'} = \exp(i\sum_i \mu_i \hat{n}_i)\ket{\psi}.$ As an effect, in the new basis we have time-independent $\Delta$ but 
the hoppings becomes $t_{ij}(t) = t_{ij}\exp(-i[\mu_i-\mu_j]t)$, since in the hoppings one needs to take into account the different reference energies in different parts of the system. We assume that the chemical potential is the same within the middle part M and the right lead R, equal to $\mu_R$, but different in the left lead $\mu_L$. Thus, $t_{LM}(t) = t_{LM}\exp(-i[\mu_L-\mu_R]t)$, similarly for the other direction, and time-independent elsewhere.

\section{Non-equilibrium Green's functions (NEGFs)}
The chemical potential bias $V$ that we impose in the two-terminal setup induces a current through the system. We calculate the current, taking into account the effects of the interaction, by using the non-equilibrium Green's functions (NEGF) method. Clear and thorough textbook treatments of the topic are given, to name a few, with standard second-quantized formalism by Refs. \cite{rammer2007,haug2008,stefanucci2013} and with path integrals by Ref. \cite{kamenev2011}. Historical roots of the method are in the works of Schwinger \cite{schwinger1961}, Kadanoff and Baym \cite{kadanoff1962} and Keldysh \cite{keldysh1965}. 
The particular approach we take to model the two-terminal setup is inspired by the 'Hamiltonian approach' to superconducting point-contact and quantum dot junctions considered in Refs. \cite{cuevas1996,martin-rodero2011}, but generalized to handle a tight-binding structure in between. 

\subsection{Green's functions and Dyson's equations for a time-periodic system}
The two-time Green's functions we are using in this work are defined in the Nambu block form
\begin{equation}
    G^{R}_{ij}(t,t') = -i\theta(t-t') 
    \begin{pmatrix}
    \meanv{[\hat{c}_{i\uparrow}(t),\hat{c}_{j\uparrow}^\dagger(t')]} & \meanv{[\hat{c}_{i\uparrow}(t),\hat{c}_{j\downarrow}(t')]} \\
    \meanv{[\hat{c}^\dagger_{i\downarrow}(t),\hat{c}_{j\uparrow}^\dagger(t')]} & \meanv{[\hat{c}^\dagger_{i\downarrow}(t),\hat{c}_{j\downarrow}(t')]}
    \end{pmatrix}
    ~,
\end{equation}
\begin{equation}
    G^{A}_{ij}(t,t') =  i\theta(t'-t) 
    \begin{pmatrix}
    \meanv{[\hat{c}_{i\uparrow}(t),\hat{c}_{j\uparrow}^\dagger(t')]} & \meanv{[\hat{c}_{i\uparrow}(t),\hat{c}_{j\downarrow}(t')]} \\
    \meanv{[\hat{c}^\dagger_{i\downarrow}(t),\hat{c}_{j\uparrow}^\dagger(t')]} & \meanv{[\hat{c}^\dagger_{i\downarrow}(t),\hat{c}_{j\downarrow}(t')]}
    \end{pmatrix}
\end{equation}
and
\begin{equation}
    G^<_{ij}(t,t') = i
    \begin{pmatrix}
        \meanv{\hat{c}^\dagger_{j\uparrow}(t')\hat{c}_{i\uparrow}(t)} & \meanv{\hat{c}_{j\downarrow}(t')\hat{c}_{i\uparrow}(t)} \\
        \meanv{\hat{c}^\dagger_{j\uparrow}(t')\hat{c}^\dagger_{i\downarrow}(t)} & \meanv{\hat{c}_{j\downarrow}(t')\hat{c}^\dagger_{i\downarrow}(t)}
    \end{pmatrix}
\end{equation}
which are known respectively as the retarded, the advanced and the lesser Green's function, respectively. It holds $G^R(t,t')^\dagger = G^A(t',t)$ and
$G^<(t,t')^\dagger = -G^<(t',t).$

% The interaction and other perturbations are given in terms of the self-energies, which are
% $\Sigma^R(t,t'), \Sigma^A(t,t'),\Sigma^<(t,t')$.
% If we denote the Green's functions for a system with $\Sigma(t,t')=0$
% by  $g^R(t,t'),g^A(t,t'),g^<(t,t')$,
% we have the Dyson's equations
% \begin{equation} 
%     G^{R/A}(t,t') = g^{R/A}(t,t') + \int \diffe{t_1}\diffe{t_2} g^{R/A}(t,t_1)\Sigma^{R/A}(t_1,t_2)G^{R/A}(t_2,t')
% \end{equation}
% and
% \begin{equation} 
%     G^{R/A}(t,t') = g^{R/A}(t,t') + \int \diffe{t_1}\diffe{t_2} G^{R/A}(t,t_1)\Sigma^{R/A}(t_1,t_2)g^{R/A}(t_2,t')
% \end{equation}
% for the retarded and advanced and the kinetic equation
% \begin{equation}
%     \begin{split}
%     G^<(t,t') &= \int\diffe{t_1}\diffe{t_2}\\
%     &\left(I\delta(t-t_1)+\int \diffe{t_3} G^R(t,t_3)\Sigma^R(t_3,t_1)\right)
%     g^<(t_1,t_2)
%     \left(I\delta(t_2-t')+\int\diffe{t_3} \Sigma^A(t_2,t_3)G^A(t_3,t')\right)\\
%     &+\int \diffe{t_1}\diffe{t_2} G^R(t,t_1)\Sigma^<(t_1,t_2)G^A(t_2,t')
%     \end{split}
% \end{equation}
% where the Green's functions and self-energies are now matrices in the Nambu basis
% and the product is the corresponding matrix product. 

Due to the time-periodicity of the Hamiltonian with the fundamental frequency $\omega_0 = V$, which is the frequency of the time-dependent hopping between the left lead and the middle system $t_{ML}(t) = t_{ML}\exp(iVt)$,
the solution of the Green's functions is simplified by using the Fourier transform on the time argument, which is defined as $F(\omega,\omega') = \int\int \diffe{t}\diffe{t'} F(t,t') \exp(i[\omega t -\omega' t'])$, while the inverse transformation is 
$F(t,t') = \frac{1}{4\pi^2}\int\int\diffe{\omega}\diffe{\omega'}F(\omega,\omega')\exp(-i[\omega t-\omega' t']).$ Since we assume that the state of the system is also time-periodic, we make the ansatz for the Green's functions (omitting the specific type since this is general) 
\begin{equation}
    G(t,t') = \sum_n G_n(t-t')\exp(in\omega_0 t)~,
    \label{eq:g_ansatz}
\end{equation}
where $G_n(t-t')$ are Fourier components, which are defined by this expression. 
Note that here we denote the total matrix corresponding to Nambu blocks $G_{ij}$ when we omit the indices. However,
the consideration is general for any two-time Green's function. 
Doing the double-time Fourier transform leads to
\begin{equation}
    G(\omega,\omega') = \sum_n G_n(\omega)2\pi \delta(\omega'-\omega-n\omega_0)~,
\end{equation}
where the time-periodicity is evident in the connection between the two frequencies.
We obtain simple forms for the Green's functions by defining the matrix components 
\begin{equation}
    G_{nm}(\omega) \equiv G_{m-n}(\omega+n\omega_0) ,
    \label{eq:harmonic_matrix_def}
\end{equation}
where now each component $nm$ corresponds to a block of $2N\times 2N$, where $N$ is the number of sites in the system. In other words,
each block corresponds to a Nambu Green's function.

By introducing the self-energies $\Sigma^<(t,t'),\Sigma^{R/A}(t,t')$ and the unperturbed Green's functions $g^{R/A}(t,t'),g^<(t,t')$ (obtained with $\Sigma=0$), we have the Dyson's equation for the retarded and advanced components in the combined harmonic and Nambu block matrix form
\begin{equation}
    G^{R/A}(\omega) = g^{R/A}(\omega) + g^{R/A}(\omega)\Sigma^{R/A}(\omega)G^{R/A}(\omega)
    =  g^{R/A}(\omega) + G^{R/A}(\omega)\Sigma^{R/A}(\omega)g^{R/A}(\omega)~,
    \label{eq:dyson}
\end{equation}
and the Kadanoff-Baym kinetic equation for the lesser component
\begin{equation}
    G^<(\omega) = \left[I+G^R(\omega)\Sigma^R(\omega)\right]g^<(\omega)\left[I+\Sigma^A(\omega)G^A(\omega)\right]
    + G^R(\omega)\Sigma^<(\omega)G^A(\omega)~.
    \label{eq:kinetic}
\end{equation}
The matrix product corresponds to summing over the intermediate Nambu and harmonic indices of the matrices. Accordingly, $I$ is the combined unit matrix in the harmonic and Nambu index.
These equations can be used to obtain the frequency components of the lesser Green's function
when the unperturbed Green's functions and the self-energy is known.

In this work, we need only the equal time lesser Green's function, which is given in terms of its Fourier series components by
\begin{equation}
    G_n^<(t,t) = \int_{0}^{\omega_0}\diffe{\omega} \mathrm{Tr}_{n} G^<(\omega)~,
    \label{eq:real_Gl_fourier}
\end{equation}
%where $\mathrm{Tr}_n$ denotes the trace over the $n$th subdiagonal of the harmonic subspace of the matrix, that is, it includes a sum over the $n$th harmonic block matrix subdiagonal. 
where $\mathrm{Tr}_n$ denotes a partial trace, where the Nambu Green's function matrices corresponding to the harmonic indices $m,m+n$ are summed over $m$, resulting in a $2N\times 2N$ matrix corresponding to $G^<_n(\omega)$.

\subsection{NEGF treatment of the two-terminal setup}
We introduce the effect of the finite bias in the calculation by
starting from the non-perturbed equilibrium problem of non-interacting particles and
with the leads disconnected from the middle system, and considering the connection to the leads and the interactions as perturbations. 
In other words, we have to amend the Hartree self-energy by the contact Hamiltonian to obtain 
$\Sigma^{R/A}(t,t') = \hat{H}_{\mathrm{contact}}(t)\delta(t-t')+\Sigma^{R/A}_{\mathrm{Hartree}}(t,t')$, where $\Sigma^{R/A}_{ij,\mathrm{Hartree}}(t,t') = -i \delta_{ij}\delta(t-t')U_i G_{ij}^<(t,t). $ Also, we have that  $\Sigma^{<}(t,t') = 0$. Here $G^<_{ij}(t,t)$ is the lesser Nambu Green's function. It is important for transport calculations that the approximation fulfills the conservation laws. This self-energy fulfills the Kadanoff-Baym criterion for fulfilling the conservation laws (and possible gauge invariance) if the Green's function $G_{ij}^<(t,t)$ is determined self-consistently \cite{baym1961,baym1962}. 

We work with the basis where superconducting order parameters in the leads are time-independent and $H_{\mathrm{contact}}(t)$ is periodically dependent on time. 
We assume in this work that the unperturbed middle part is in equilibrium with the right lead, that is, they have the same chemical potential. In other words, the bias is between the left lead and the middle part. Thus, the time-period  corresponds to the fundamental frequency $\omega_0 = V$.

Now that we have the equations of motion, that is, the Dyson's equations \eqref{eq:dyson}, \eqref{eq:kinetic}
and we have a way to determine the self-energy $\Sigma^{R/A}$, the remaining task is to determine the non-perturbed Green's functions.
Since the non-perturbed system is assumed to be at equilibrium, an important tool for the purpose is the fluctuation-dissipation theorem, which states that, for fermions at the zero chemical potential $\mu=0$,
\begin{equation}
    g_{nn}^<(\omega) = f_{FD}(\omega+n\omega_0)(g^A_{nn}(\omega)-g^R_{nn}(\omega))~,
\end{equation}
where $f_{FD}$ is the Fermi-Dirac distribution. This result can be understood based on the fact that $A(\omega) = g^A(\omega)-g^R(\omega)$ is the spectral function, which tells the density of states of the system and $g^<(\omega)$ tells the observable density expectation values: at equilibrium the states are filled according to Fermi-Dirac statistics, giving the densities. Using the fluctuation-dissipation theorem, it is sufficient to determine the retarded and advanced components in order to obtain the lesser Green's function. Note that at equilibrium, the non-diagonal components (in the harmonic index $n$) of the Green's functions are zero.

For the middle system, we have
\begin{equation}
    g^{R/A}_{MM,nn}(\omega) = \left((\omega+n\omega_0 \pm i\eta)I-T_{MM}\right)^{-1}~,
\end{equation}
where $T_{MM}$ is the tight-binding matrix of the middle part written in the Nambu basis and $\pm i\eta$ is a small regularization parameter to ensure 
convergence to the proper Green's function in the inverse Fourier transform.
For the superconducting leads we assume the wide-band limit where we have \cite{cuevas1996}
\begin{equation}
    \begin{split}
    g^{R/A}_{LL/RR,nn}(\omega)
    &= \frac{1}{t_{L/R}\sqrt{\abso{\Delta_{L/R}}^2-(\omega+n\omega_0\pm i\eta)^2}} \\
    &\begin{pmatrix}
        -(\omega+n\omega_0\pm i\eta) & -\Delta_{L/R} \\
        -\Delta_{L/R}^* & -(\omega+n\omega_0\pm i\eta) \\
    \end{pmatrix}~.
    \end{split}
\end{equation}
It can be checked that this has the typical density of states.
For the normal leads, which we consider as linear chains, we have
\begin{equation}
    g^{R/A}_{LL/RR,nn}(\omega)
    =\frac{1}{2t_{L/R}}
    \begin{pmatrix}
        \exp(\mp i\phi) & 0 \\
        0 & \exp(\mp i\phi) \\
    \end{pmatrix}~,
\end{equation}
where $\phi = \arccos(\omega+n\omega_0)$.
Note that the fluctuation-dissipation theorem at a finite chemical potential would give the lesser Green's function as
\begin{equation}
    g^{<}_{LL/RR,nn}(\omega)
    =\frac{1}{t_{L/R}}
    \begin{pmatrix}
        f_{FD}(\omega+n\omega_0-\mu_{L/R})\sin(\phi) & 0 \\
        0 & f_{FD}(\omega+n\omega_0+\mu_{L/R})\sin(\phi) \\
    \end{pmatrix}~,
\end{equation}
where we note that the particles and holes have the chemical potential with opposite signs.
In the wide-band limit $t_{L/R} \gg \omega+n\omega_0$, where $\omega+n\omega_0$ includes the energies of interest,
the Green's function is given by $g^{R/A}_{LL/RR,nn}(\omega) = \mp i \frac{1}{t_{L/R}}$. The normal lead Green's function formulas, especially for the finite chemical potential, are important if one wants to formulate the $NN$ and $NS$ junctions in the picture where the lead-system hoppings are time-independent and the chemical potentials appear as usual.

Finally, we need the observables of interest in terms of the Green's functions. For self-consistent determination, we require the particle number and the superconducting order parameter,
which are given by
\begin{equation}
    \meanv{n_{i\uparrow}(t)} = \mathrm{Im}(G^<_{0,i\uparrow,i\uparrow}) +
    \sum_{n=1}^\infty \left( 2\mathrm{Im}(G^<_{n,i\uparrow,i\uparrow})\cos(nVt)
    +2 \mathrm{Re}(G^<_{n,i\uparrow,i\uparrow})\sin(nVt)\right)~,
\end{equation}
which also gives the down spin particle number due to our assumption assumption $\meanv{n_{i\uparrow}} = \meanv{n_{i\downarrow}}$, and
\begin{equation}
    \meanv{\Delta_i(t)} 
    = \sum_{n=-\infty}^\infty 
    -iG^<_{n,i\uparrow,i\downarrow}\exp(inVt)~.
\end{equation}
Assuming that the hopping $t_{ij}(t)$ between sites $i$ and $j$ oscillates in time with the frequency $p\omega_0$, that is, its Fourier component $t_{p,ij}$ is finite others being zero, the current is given by
\begin{equation}
\begin{split}
    I_{ij}(t) = -4\mathrm{Re}(t_{p,ij}G^<_{-p,j\uparrow,i\uparrow})
    -  \sum_{n=1}^{\infty} \bigg(&4\mathrm{Re}(t_{p,ij}G^<_{n-p,j\uparrow,i\uparrow}-t_{-p,ji}G^<_{n+p,i\uparrow,j\uparrow})\cos(nVt) \\
    &-4\mathrm{Im}(t_{p,ij}G^<_{n-p,j\uparrow,i\uparrow}-t_{-p,ji}G^<_{n+p,j\uparrow,i\uparrow})\sin(nVt)\bigg)~.
    \end{split}
\end{equation}

\subsection{Numerical details}
We obtain the real-time Green's functions from the solved non-equilibrium Green's functions in Fourier space by using the inverse Fourier transform formula Eq. \eqref{eq:real_Gl_fourier} and the Fourier series ansatz Eq. \eqref{eq:g_ansatz}. Firstly, we note that the harmonic-Nambu basis is, in principle, infinite due to the infinite size of the system and the infinite amount of harmonics of the basic frequency $\omega_0 = V$.
In order to make the harmonic part numerically tractable, we introduce frequency cutoffs from above and below, $\omega_u,\omega_d,$ respectively.  The number of the harmonic block indices is then $\omega_u-\omega_d)/V$. The cutoffs have to be chosen so that the states in the middle structure are within them. Furthermore, the regularization parameter $i\eta$ in the retarded and advanced Green's functions spread the spectral densities. An important thing to note is that with the SS junction at the zero temperature, one cannot cut the energies above the Fermi energies: the MAR couple to frequencies up to the superconducting order parameter and also slightly above since the AR probability does not go to zero immediately above gap. One has to test numerically, which are suitable limits. We found that the interval $[-8|t_{AA}|,3|t_{AA}|]$ was suitable for the sawtooth lattice. Its total bandwidth is around $6|t_{AA}|$.

Also, the system itself is infinite and therefore the Green's functions in the single-particle basis would be infinite. However, it turns out we do not need to solve for all the sites of the system explicitly. By looking at the Dyson's equations, it is clear that a closed system of equations are formed for the Green's functions $G_{ij}$ with indices $i,j$ corresponding to finite self-energies in the sense that $\Sigma_{\cdot,i},\Sigma_{i,\cdot}, \Sigma_{\cdot,j},$ or $\Sigma_{j,\cdot}$, where $\cdot$  corresponds to any index, is finite. Other Green's functions can be calculated from this closed set, if needed. Thus, in order to limit the size of the Nambu site basis, we consider the subsystem with the contact sites at the leads and the middle part.   

We note that in general the integrand in the inverse Fourier transform, needed to obtain the real-time Green's functions, can be strongly peaked around certain values of frequency, especially in the case of flat bands. Thus, standard trapezoid rule is not sufficient for determining the integral accurately with reasonable effort. In order to have an accurate determination of the integral with relatively few function evaluations, which are expensive for large systems, we utilize the doubly adaptive quadrature algorithm \cite{espelid2003,espelid2007} to evaluate the integral. The relative accuracy of $10^{-7}$ is demanded in the numerical calculations, which is usually achieved with evaluating the frequency Green's function at a couple of hundred to a thousand frequency values. The frequency cutoffs were introduced to allow computations at finite times. However, we tested that they do not introduce qualitative difference.

We solve the problem self-consistently by the standard iteration techniques but including, on top of the time-averaged components, the finite harmonics of the base frequency $\omega_0$ of the Hartree potential and the superconducting order parameter up to a cutoff. According to our knowledge, the alternative method for getting the self-consistent solution based on minizing a functional, e.g. free energy, is not available at non-equilibrium conditions. We note that the direct fixed-point iteration does not always converge. We utilize mixing, in the particular the Anderson's method and Broyden's method \cite{anderson1965,pulay1982,broyden1965} to allow and boost the covergence of the algorithm. In the numerical calculations shown in this work, we determined the sufficient number of Fourier components until the higher contributions were found to be lesser than $10^{-4}$ relative to the largest components. This demands keeping at most four finite harmonics. The self-consistency was tested based on the maximum metric, that is, taking the maximum component-wise relative error in the self-consistent parameters. Accuracy of $10^{-5}$ was required for the solutions. 

We note that for the case of flat bands, obtaining the solution is usually difficult due to the Hartree potential varying quickly with the gate potential that controls the filling. Thus, small error in the Hartree potential can introduce a large difference in the filling, which makes the determination less stable against numerical error than is usually the case. However, we have been able to obtain self-consistency.

In this work, we assume that the bias drops at the junction between the middle part and the left lead. 
If one looks at the Josephson voltage-phase relation $\deriv{\phi}{t} = 2V$, it follows that the AC Josephson current is finite where the chemical potential varies. In the time-periodic ansatz we have, we assume that all the sites at the middle part vary in time in harmonics of $\omega_0 = 2V$. This does not allow for a more continuous distribution of voltage drop in the junction with a superconducting junction. In the normal state, this would not matter. 
%However, with finite superconducting order parameter $\Delta$ at the junction, for instance voltage drops of $V/N$ over $N$ sites would allow, by the Josephson voltage-phase relation, relative phase-oscillation with frequency $\omega_0' = 2V/N$ between neighboring sites. Note that non-commensurate general non-uniform voltage drop would not lead to a stationary state. 
Assuming this is the case also for superconductors, we have put the voltage drop at the edge since a more detailed approach would lead to too large computation times while keeping the algorithm otherwise the same. The effect of this approximation is to focus the AC Josephson effect to the edge.